\documentclass{article} 
\PassOptionsToPackage{table}{xcolor}
\usepackage{iclr2027_conference,times}

\usepackage{amsmath,amsfonts,bm}

\def\eqref#1{equation~\ref{#1}}

\def\1{\bm{1}}

\DeclareMathAlphabet{\mathsfit}{\encodingdefault}{\sfdefault}{m}{sl}
\SetMathAlphabet{\mathsfit}{bold}{\encodingdefault}{\sfdefault}{bx}{n}

\usepackage{hyperref}
\usepackage{url}
\usepackage{booktabs}
\usepackage{threeparttable}
\usepackage{multirow}
\usepackage{amsmath}
\usepackage{graphicx}
\usepackage{subcaption}
\usepackage{siunitx}
\usepackage{pifont}
\usepackage{tikz}
\usepackage{helvet}
\usepackage{enumitem}
\usepackage{stfloats}
\fnbelowfloat
\newlist{claims}{itemize}{1}
\setlist[claims]{leftmargin=1.25em,itemsep=1.5pt,topsep=3pt,parsep=0pt,label=\textbullet}
\usetikzlibrary{positioning,arrows.meta,fit,backgrounds,shapes.geometric,calc}
\definecolor{iorange}{RGB}{153,0,0}
\definecolor{icyan}{RGB}{96,142,180}
\definecolor{iblue}{RGB}{255,204,0}
\definecolor{igray}{RGB}{142,142,147}
\definecolor{igreen}{RGB}{0,105,120}
\newcommand{\ci}[1]{{\scriptsize\textcolor{igray}{$\pm$\,#1}}}
\title{\textsc{MyoCodec}: A Streaming Neural Codec for Electromyography}

\author{%
\parbox[t]{\dimexpr\textwidth-2\tabcolsep\relax}{\centering
Jihwan Lee$^{1}$\quad Kleanthis Avramidis$^{1}$\quad Junhyeok Lee$^{2}$\quad Tiantian Feng$^{1}$\\
Najim Dehak$^{2}$\quad Shrikanth Narayanan$^{1}$\\[4pt]
\normalfont
$^{1}$Signal Analysis and Interpretation Lab (SAIL), University of Southern California, USA\\
$^{2}$Center for Language and Speech Processing, Johns Hopkins University, USA\\
\texttt{jihwan@usc.edu}}%
}

\iclrfinalcopy  

\begin{document}

\maketitle

\lhead{Preprint.}

\vspace{-5mm}

\begin{abstract}

\vspace{-3mm}

Neural codecs encode continuous signals into compact sequences of discrete tokens, providing an interface for efficient transmission, storage, and token-based sequence modeling.
This paradigm has been widely adopted in modern speech and audio frameworks; however, the biosignal domain still lacks a neural codec designed specifically for low-bitrate streaming and generalization across diverse downstream tasks.
\textbf{We present \textsc{MyoCodec}, a streaming neural codec designed for electromyography (EMG)}. Inspired by recent neural audio codecs, \textsc{MyoCodec} combines causal Transformers with residual vector quantization to encode continuous EMG signals into different levels of EMG representations spanning from continuous latent features to discrete tokens operating at \SI{50}{Hz}. Trained on twelve public EMG datasets, \textsc{MyoCodec} achieves favorable performance in both intrinsic codec quality and representative downstream tasks, including typing (\textit{emg2qwerty}), hand-pose (\textit{emg2pose}), speech decoding (\textit{emg2speech}), and speech-to-EMG synthesis (\textit{speech2emg}). Across these tasks, \textsc{MyoCodec} exhibits strong performance against prior models while providing a compact and causal EMG representation.
During streaming inference, it requires compute time of only \SI{0.482}{ms} for each \SI{20}{ms} frame, enabling real-time streaming.
Also, the discrete token representation provided by \textsc{MyoCodec} has the potential to support integration into language-model based approaches, creating a path toward LLM-based interactive systems, where tokenized EMG representations are directly processed into such language or speech models. Code and model weights are released\footnote{\url{https://github.com/lee-jhwn/myocodec}}.

\end{abstract}

\section{Introduction}

Foundation models have reshaped various domains, such as text, speech, audio, and biosignal, by replacing task-specific front ends with generalizable representations learned from large-scale unlabeled data. Rather than training a separate encoder for every application, a single pretrained model can provide a shareable representation for various tasks, including recognition, generation, and understanding across a broad range of downstream tasks.

In the biosignal domain, LaBraM learns general-purpose representations for EEG \citep{jiang2024labram}, while the generic neuromotor interface of \citet{kaifosh2025neuromotor} demonstrates that EMG decoding can generalize across individuals without per-user calibration. However, such approaches in the EMG domain still lack a shared, task-agnostic representation for diverse downstream tasks. To serve as a practical interface in the EMG domain, such a representation must not only transfer across tasks but also be compact, streamable, and lightweight.

In the speech and audio domain, these requirements have been met through the neural codec approach.
Neural audio codecs such as SoundStream \citep{zeghidour2021soundstream}, EnCodec \citep{defossez2022encodec}, DAC \citep{kumar2023dac}, Mimi \citep{defossez2024moshi}, and JHCodec \citep{lee2026ssrr} combine an encoder, residual vector quantization (RVQ), and a decoder trained with reconstruction and adversarial objectives to encode continuous waveforms into compact discrete tokens. The resulting representation can serve several purposes, such as reducing transmission bandwidth, limiting storage and caching costs, and providing tokens that can be modeled directly by modern token-based models. These advantages position neural codecs as general-purpose front ends for speech and audio systems.

BioCodec \citep{avramidis2025biocodec} is one of the first to extend this neural-codec paradigm to electrophysiology signals. It adapts the EnCodec~\citep{defossez2022encodec} architecture using causal convolutions and RVQ, producing discrete representations that generalize across various downstream tasks, such as clinical, sleep, speech, and motor-imagery tasks. BioCodec establishes the central premise that the neural-codec paradigm can successfully be applied to biosignals.

However, BioCodec has two main limitations. First, it focuses primarily on electroencephalography (EEG), with EMG treated only as a secondary modality. 
Second, despite its causal architecture, its experimental design and implementation lack a frame-wise streaming aspect.
These limitations lead to an open question for on-device EMG frameworks: whether a codec designed with the temporal and deployment constraints of EMG can produce low-bitrate discrete representations in real time. For more details of related works, refer to Appendix~\ref{app:related}.

Bridging this gap, \textbf{we present \textsc{MyoCodec}, a streaming neural codec designed for EMG signals}. \textsc{MyoCodec} consists of an encoder, a residual vector quantizer, and a decoder. The encoder and decoder are based on causal Transformers to support streamability. \textsc{MyoCodec} takes \SI{2}{kHz} EMG signals as input and encodes them into different levels of EMG representations spanning from continuous encoder features to discrete token streams operating at \SI{50}{Hz} and \SI{2400}{bits/s} per channel. \textsc{MyoCodec} is trained in two phases. First, it is trained with signal reconstruction objectives and then with a gradually introduced adversarial objective. It is trained on twelve EMG datasets that are publicly available. On a single \texttt{NVIDIA RTX PRO 6000 Blackwell} GPU, the inference time takes \SI{0.48}{ms} per frame, achieving a real-time throughput of $41\times$.

We probe \textsc{MyoCodec} in two aspects: intrinsic codec quality and generalizability to downstream tasks. Experimental results show that \textsc{MyoCodec} outperforms existing EMG representation models in signal reconstruction.
\textsc{MyoCodec} also provides useful EMG representations generalizing to various downstream tasks, including typing (\textit{emg2qwerty}~\citep{sivakumar2024emg2qwerty}), hand-pose detection (\textit{emg2pose}~\citep{salter2024emg2pose}), speech decoding (\textit{emg2speech}~\citep{gaddy2021improved}), and EMG synthesis (\textit{speech2emg}~\citep{scheck2023stegan}).
Our experimental results demonstrate that \textsc{MyoCodec} outperforms or closely matches existing frameworks, even in a streaming setting, showing its generalization to various applications.


Beyond its current performance, \textsc{MyoCodec} provides a future potential integration between muscle activity signals and modern token-based sequence modeling approaches. 
By encoding continuous EMG into low-bitrate discrete tokens, it enables training objectives widely used in modern language modeling approaches, such as categorical next-token prediction. Neural audio codecs have similarly enabled Transformer-based modeling of speech and audio in systems such as AudioLM, VALL-E, and Moshi \citep{borsos2023audiolm,wang2023valle,defossez2024moshi}, leading to substantial improvement in performance.
Similarly, \textsc{MyoCodec} tokens could further be incorporated into language or speech-language models, enabling unified systems that map muscle activity directly to language, speech, or actions. Such systems could reduce reliance on cascaded pipelines, in which latency and errors accumulate across separately trained stages.

Other potential applications include silent-speech communication, hands-free text entry, gesture-based interaction, prosthetic control, and other assistive or wearable interfaces. Also, the ability of \textsc{MyoCodec} to generate EMG signals could support synthetic biosignal augmentation, simulation of task-conditioned muscle activity, and personalized neuromuscular digital twins. Moreover, the \textsc{MyoCodec}'s lightweight model size, streamability, and fast inference speed also make on-device processing plausible, reducing communication bandwidth and limiting the transmission of sensitive raw biosignals.

Our main contributions include:
\vspace{-3mm}
\begin{claims}
\item We introduce \textsc{MyoCodec}, a streaming neural codec for EMG, operating at \SI{50}{Hz} and \SI{2400}{bits/s} per channel.

\item We demonstrate its strong intrinsic codec quality and transferability across diverse downstream tasks, establishing \textsc{MyoCodec} as a generalizable EMG representation.

\item With lightweight model size and fast streaming inference, \textsc{MyoCodec} provides a practical framework for low-latency wearable systems and future on-device deployment.

\end{claims}

\section{Method}

\label{sec:method}

\subsection{Architecture}
\label{sec:arch}

\begin{figure}[t]
\centering
\resizebox{\textwidth}{!}{%
\begin{tikzpicture}[
  x=1cm,y=1cm,font=\sffamily\bfseries\small,
  >={Latex[length=1.8mm]},
  ar/.style={->,line width=0.7pt},
  cont/.style={fill=igray!20,rounded corners=4pt,inner sep=0pt},
  qcont/.style={fill=icyan!10,rounded corners=4pt,inner sep=0pt},
  blk/.style={draw=black,line width=0.7pt,rounded corners=2.5pt,fill=icyan!30,
              minimum width=6.4mm,minimum height=25mm,inner sep=0pt,align=center},
  tfb/.style={draw=black,line width=0.7pt,rounded corners=2.5pt,fill=iorange!40,
              minimum width=6.4mm,minimum height=25mm,inner sep=0pt,align=center},
  vq/.style={draw=iblue!80!black,line width=0.7pt,rounded corners=2.5pt,fill=iblue!40,
             minimum width=5.6mm,minimum height=13.5mm,inner sep=0pt,align=center},
  op/.style={circle,draw=black,line width=0.7pt,fill=black,text=white,
             inner sep=0pt,minimum size=4.2mm,font=\sffamily\bfseries\scriptsize},
  loss/.style={draw=black,line width=0.7pt,rounded corners=3pt,fill=iorange,
               text=white,align=center,inner sep=3pt,font=\sffamily\bfseries\small,
               minimum width=44mm,minimum height=9mm},
  lar/.style={->,draw=iorange,line width=0.7pt},
  dar/.style={->,draw=igreen,line width=0.7pt,dash pattern=on 2.4pt off 1.8pt},
  dim/.style={font=\sffamily\footnotesize,inner sep=1pt},
  lab/.style={font=\sffamily\bfseries\small},
  wave/.style={fill=black,inner sep=0pt},
]
\def\wave#1{%
  \foreach \w/\y in {0.16/0.80, 0.24/0.70, 0.12/0.60, 0.32/0.50, 0.08/0.40,
                     0.16/0.30, 0.32/0.20, 0.56/0.10, 0.80/0.00, 0.40/-0.10,
                     0.64/-0.20, 0.16/-0.30, 0.48/-0.40, 0.16/-0.50,
                     0.32/-0.60, 0.16/-0.70, 0.08/-0.80}
    \node[wave,minimum width=\w cm,minimum height=0.07cm] at ($#1+(0,\y)$) {};}
\coordinate (wl) at (0.50,0);   \wave{(wl)}
\coordinate (wr) at (15.92,0);  \wave{(wr)}
\node[lab] at (0.50,-1.16) {Waveform};
\node[lab,align=center] at (15.92,-1.30) {Reconstructed\\Waveform};

\begin{scope}[on background layer]
  \node[cont,fit={(1.38,-1.45) (5.22,1.45)}] (encbox) {};
\end{scope}
\node[blk] (e1) at (1.82,0) {\rotatebox{270}{Reshape}};
\node[blk] (e2) at (2.70,0) {\rotatebox{270}{Linear}};
\node[tfb] (e3) at (3.62,0) {\rotatebox{270}{Transformer $\times 8$}};
\node[blk] (e4) at (4.62,0) {\rotatebox{270}{Linear}};
\node[lab,below=1.6mm of encbox] {Encoder};
\foreach \a/\b in {e1/e2,e2/e3,e3/e4} \draw[ar] (\a) -- (\b);
\draw[ar] (1.02,0) -- (e1);

\begin{scope}[on background layer]
  \node[qcont,fit={(5.42,-1.92) (10.78,2.18)}] (qbox) {};
\end{scope}
\node[lab] at (7.95,1.82) {Residual Vector Quantizer};
\node[dim] at (5.76,0.32) {$\mathbf{z}_e$};
\node[vq] (v1) at (6.42,0) {\rotatebox{270}{VQ 1}};
\node[vq] (v2) at (7.58,0) {\rotatebox{270}{VQ 2}};
\node[vq] (vk) at (9.95,0) {\rotatebox{270}{VQ $K$}};
\node at (8.70,0)     {$\cdots$};
\node at (8.70,1.15)  {$\cdots$};
\node at (8.70,-1.15) {$\cdots$};
\node[op] (m1) at (7.00,1.15) {$-$};
\node[op] (m2) at (8.16,1.15) {$-$};
\node[op] (mk) at (9.25,1.15) {$-$};
\node[op] (p1) at (7.00,-1.15) {$+$};
\node[op] (p2) at (8.16,-1.15) {$+$};
\node[op] (pk) at (10.30,-1.15) {$+$};
\draw[ar] (5.22,0) -- (6.12,0);
\draw[ar] (5.90,0) |- (m1.west);
\draw[ar] (v1.east) -- (7.00,0) -- (m1.south);
\draw[ar] (m1.east) -- (7.58,1.15);
\draw[ar] (7.58,1.15) -- (v2.north);
\draw[ar] (7.58,1.15) -- (m2.west);
\draw[ar] (v2.east) -- (8.16,0) -- (m2.south);
\draw[ar] (m2.east) -- (8.48,1.15);
\draw[ar] (8.92,1.15) -- (mk.west);
\draw[ar] (mk.east) -- (9.95,1.15);
\draw[ar] (9.95,1.15) -- (vk.north);
\draw[dotted,line width=0.7pt] (5.90,-1.15) -- (p1.west);
\draw[ar] (v1.east) -- (7.00,0) -- (p1.north);
\draw[ar] (v2.east) -- (8.16,0) -- (p2.north);
\draw[ar] (vk.east) -- (10.30,0) -- (pk.north);
\draw[ar] (p1.east) -- node[dim,above,inner sep=1.5pt] {$\mathbf{z}_1$} (p2.west);
\draw[ar] (p2.east) -- (8.48,-1.15);
\draw[ar] (8.92,-1.15) -- node[dim,above,inner sep=1.5pt,pos=0.62] {$\mathbf{z}_{K-1}$} (pk.west);
\draw[ar] (pk.east) -- (10.65,-1.15) -- (10.65,0)
          -- node[dim,above,inner sep=1.5pt] {$\mathbf{z}_K$} (11.25,0);

\begin{scope}[on background layer]
  \node[cont,fit={(11.25,-1.45) (14.77,1.45)}] (decbox) {};
\end{scope}
\node[blk] (d1) at (11.67,0) {\rotatebox{270}{Linear}};
\node[tfb] (d2) at (12.59,0) {\rotatebox{270}{Transformer $\times 8$}};
\node[blk] (d3) at (13.51,0) {\rotatebox{270}{Linear}};
\node[blk] (d4) at (14.39,0) {\rotatebox{270}{Reshape}};
\node[lab,below=1.6mm of decbox] {Decoder};
\foreach \a/\b in {d1/d2,d2/d3,d3/d4} \draw[ar] (\a) -- (\b);
\draw[ar] (d4) -- (15.45,0);

\node[loss,align=center] (vqloss) at (2.90,2.76) {VQ Loss $+$ Commit Loss};
\draw[lar] (6.42,2.18) -- (6.42,2.76) -- (vqloss.east);

\node[loss] (huber) at (7.30,-3.20) {Huber Recon Loss};
\node[loss] (stft)  at (7.30,-4.35) {Spectral Recon Loss};
\draw[lar] (0.50,-1.48) -- (0.50,-3.20) -- (huber.west);
\draw[lar] (0.50,-1.48) -- (0.50,-4.35) -- (stft.west);
\draw[lar] (15.92,-1.78) -- (15.92,-2.50) -- (10.30,-2.50) -- (10.30,-3.20) -- (huber.east);
\draw[lar] (10.30,-2.50) -- (10.30,-4.35) -- (stft.east);

\draw[draw=black,line width=0.7pt,fill=igreen]
     (11.30,-4.56) -- (11.30,-3.00) -- (15.40,-3.40) -- (15.40,-4.16) -- cycle;
\node[text=white,align=center,font=\sffamily\bfseries\small] at (13.35,-3.78)
     {Discriminators\\{\sffamily\mdseries\scriptsize (stage 2)}};
\draw[dar] (16.08,-1.78) -- (16.08,-3.78) -- (15.40,-3.78);
\draw[dar] (0.34,-1.48) -- (0.34,-5.40) -- (13.35,-5.40) -- (13.35,-4.37);
\end{tikzpicture}}
\caption{\textbf{\textsc{MyoCodec} architecture.} A causal Transformer encoder--decoder with residual vector quantization (RVQ) processes each EMG channel independently at \SI{50}{Hz}. The discriminators are used only during the second phase of training.}
\label{fig:arch}
\end{figure}
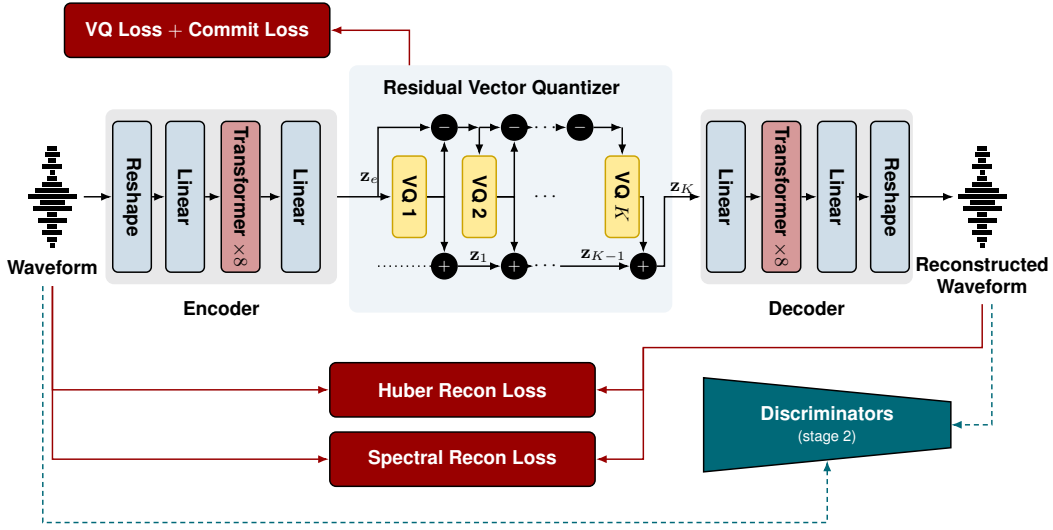

We base our architecture on the recent neural audio codec JHCodec \citep{lee2026ssrr} and adopt the training objectives from BioCodec \citep{avramidis2025biocodec}. As illustrated in Figure~\ref{fig:arch}, our model follows the standard encoder--quantizer--decoder structure in neural codec approaches. 
To support online inference, the entire architecture is causal. For more details on the architecture, refer to Appendix~\ref{app:codec}.

\paragraph{Encoder.}
The encoder takes continuous EMG signals at \SI{2}{kHz} as input.
Following BioCodec, each channel is treated independently.
Each frame consists of patchified \(40\) samples, resulting in a frame rate of \(\SI{50}{Hz}\).
A linear layer projects each frame to match the model dimension.
The projected sequence is then processed by an eight-layer Transformer with eight attention heads.
We use the rotary positional embeddings \citep{su2021roformer} and apply causal self-attention with a sliding window of 64 frames per layer.
During streaming inference, each layer maintains a 64-frame key--value cache.

\paragraph{Quantizer.}
The encoder output is projected to 32 dimensions and quantized using a residual vector quantizer with 6 stages, each stage containing a codebook of 256 entries.
A linear projection follows to map the quantized representation back to the Transformer dimension before the decoder.
To prevent codebook collapse, entries whose exponentially averaged usage falls below a fixed threshold are reinitialized from active encoder outputs every 250 optimization steps.
This procedure maintains full codebook utilization and prevents a failure mode in which the vector-quantization loss approaches zero while reconstruction quality deteriorates.

\paragraph{Decoder.}
The decoder mirrors the encoder, using an eight-layer causal Transformer with the same configuration. A final linear projection maps each decoder output back to a frame of 40 samples, resulting in reconstructed EMG signals at \SI{2}{kHz}. The encoder and decoder contain approximately \(6.3\) million parameters each, for a total model size of \(12.72\) million parameters.

\paragraph{Discriminators.}
In the second phase of training, we apply adversarial training using two discriminators: a multi-period discriminator and a multi-scale STFT discriminator.

\subsection{Training Curriculum}
\label{sec:objective}

We train \textsc{MyoCodec} in two phases.
Phase~1 jointly optimizes signal reconstruction and
vector-quantization objectives without adversarial training. To further refine reconstruction fidelity, Phase~2 adds
adversarial and feature-matching objectives while retaining the Phase~1 objectives. More details are in Appendix~\ref{app:codec}.

\paragraph{Phase~1: Reconstruction and quantization training.}
In Phase~1, we focus on joint training of vector quantization and signal reconstruction, encouraging stable, well-utilized codebooks. 
Following BioCodec, we use multiple training objectives including time-domain reconstruction $\mathcal{L}_{\text{Huber}}$, multi-scale spectral reconstruction $\mathcal{L}_{\text{spec}}$, and vector-quantization $\mathcal{L}_{\text{VQ}}$, and commit $\mathcal{L}_{\text{commit}}$ losses:
\begin{equation}
\mathcal{L}_{\text{Phase1}} = \mathcal{L}_{\text{Huber}}(x, \hat{x})
+ \mathcal{L}_{\text{VQ}}
+ \lambda_c \mathcal{L}_{\text{commit}}
+ \frac{\lambda_s}{3} \sum_{n \in \{64,128,256\}} \mathcal{L}_{\text{spec}}^{(n)}(x, \hat{x}),
\label{eq:gen}
\end{equation}
where \(x\) and \(\hat{x}\) denote the input and reconstructed signals.
We set \(\lambda_s=1\) and \(\lambda_c=0.25\).
For an STFT \(S^{(n)}\) with FFT size \(n\), the spectral loss $\mathcal{L}_{\text{spec}}^{(n)}$ is defined as:
\begin{equation}
\mathcal{L}_{\text{spec}}^{(n)} =
\big\|\log|S^{(n)}_{x}| - \log|S^{(n)}_{\hat{x}}|\big\|_1
+ \frac{\big\||S^{(n)}_{x}| - |S^{(n)}_{\hat{x}}|\big\|_\mathcal{F}}{\big\||S^{(n)}_{x}|\big\|_\mathcal{F}}
+ 0.1 \big\|\angle S^{(n)}_{x} - \angle S^{(n)}_{\hat{x}}\big\|_1,
\label{eq:spec}
\end{equation}
where $\|\cdot\|_{\mathrm{\mathcal{F}}}$ indicates the Frobenius norm. Each component captures log-magnitude error, spectral convergence, and phase error, respectively.
We train Phase~1 for 100k steps using the Muon optimizer~\citep{jordan2024muon} with a batch size of 1024.

\paragraph{Phase~2: Adversarial training.}
In Phase~2 training, we introduce gradual adversarial training to improve the fine-grained quality of the reconstructed signals while retaining the Phase~1 objectives:
\begin{equation}
\mathcal{L}_{\mathrm{Phase 2}}
=
\mathcal{L}_{\mathrm{Phase 1}}
+
\lambda_{\mathrm{adv}}\mathcal{L}_{\mathrm{adv}}
+
\lambda_{\mathrm{FM}}\mathcal{L}_{\mathrm{FM}} .
\label{eq:adv}
\end{equation}
where $\mathcal{L}_{\mathrm{adv}}$ and $\mathcal{L}_{\mathrm{FM}}$ denote the adversarial and feature-matching loss, respectively.
Their weights are increased linearly from zero to $\lambda_{\mathrm{adv}}=0.1$ and $\lambda_{\mathrm{FM}}=2.0$ over the first 50k steps of Phase~2, providing a stable transition to adversarial training. We train Phase~2 for an additional 100k steps with a batch size of 192.

\begin{table}[tb]
\centering
\begin{threeparttable}
\caption{\textbf{Intrinsic codec quality.} Reconstruction quality is evaluated using
SI-SDR and SNR (dB), waveform Pearson correlation \(r_{\mathrm{wav}}\), and envelope
Pearson correlation \(r_{\mathrm{env}}\). \(N_q\) is the number of codebooks used. \textsc{MyoCodec} outperforms TinyMyo and BioCodec in signal reconstruction by using only two codebooks (\SI{800}{bits/s}).}
\label{tab:recon}
\small
\setlength{\tabcolsep}{3.5pt}
\begin{tabular}{lrrrrrrrrrr}
\toprule
\multirow{2}{*}{Model} & \multirow{2}{*}{Params} & \multirow{2}{*}{Hz} & \multirow{2}{*}{\(N_q\)}
  & \multirow{2}{*}{bitrate/ch} & \multicolumn{4}{c}{reconstruction $\uparrow$}
  & \multicolumn{2}{c}{codebook} \\
\cmidrule(lr){6-9}\cmidrule(lr){10-11}
& & & & & SI-SDR & SNR & \(r_{\mathrm{wav}}\) & \(r_{\mathrm{env}}\) & util. & perp. \\
\midrule
TinyMyo & \SI{3.6}{M} & 100.0 & --- & $\le$614k\tnote{$\dagger$} & 0.46 & 1.70 & 0.735 & 0.445 & --- & --- \\
BioCodec & \SI{12.3}{M} & 27.8 & 6 & 1333 & $-1.51$ & 2.20 & 0.643 & 0.907 & 1.000 & 203.4 \\
\midrule
\multirow{3}{*}{\textsc{MyoCodec} (Ours)} & \multirow{3}{*}{\SI{12.7}{M}} & \multirow{3}{*}{50.0}
  & 2 & 800 & 3.62 & 4.99 & 0.832 & 0.933 & 1.000 & 172.1 \\
 & & & 3 & 1200 & 5.89 & 6.75 & 0.881 & 0.949 & 1.000 & 177.8 \\
 & & & 6 & 2400 & \textbf{9.80} & \textbf{9.98} & \textbf{0.933} & \textbf{0.967} & 1.000 & 186.0 \\
\bottomrule
\end{tabular}
\begin{tablenotes}[flushleft]\footnotesize
\item[$\dagger$] Rate assumes \texttt{float32}.
\end{tablenotes}
\end{threeparttable}
\end{table}

\begin{figure}[t!]
\centering
\includegraphics[width=\textwidth]{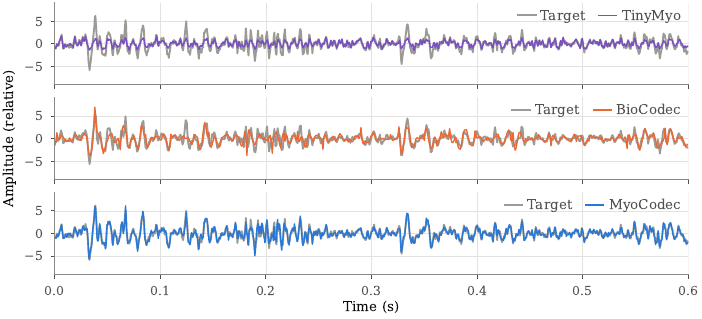}
\caption{An example of EMG signals reconstructed by each model: TinyMyo (top), BioCodec (middle), and \textsc{MyoCodec} (bottom). \textsc{MyoCodec} shows the best reconstruction quality, more properly capturing signal peaks missed by the other models.}
\label{fig:recon}
\end{figure}

\begin{table}[tb]
\centering
\caption{Per-codebook utilization and perplexity for \textsc{MyoCodec}. All six codebooks are fully utilized, with higher perplexity in later RVQ stages.}
\label{tab:cbusage}
\small
\begin{tabular}{lrrrrrrr}
\toprule
 & CB 1 & CB 2 & CB 3 & CB 4 & CB 5 & CB 6 & mean \\
\midrule
Utilization & 1.000 & 1.000 & 1.000 & 1.000 & 1.000 & 1.000 & 1.000 \\
Perplexity & 154.9 & 189.2 & 189.2 & 192.6 & 194.1 & 196.0 & 186.0 \\
\bottomrule
\end{tabular}
\end{table}

\subsection{Training Data and Preprocessing}
\label{sec:data}

We train \textsc{MyoCodec} on twelve publicly available EMG datasets. They contain approximately 24k channel-hours of data from 250k recordings. We use the official training, validation, and test set partitions of each dataset when available and construct speaker- and/or session-disjoint random splits otherwise. For datasets subsequently used in downstream evaluation, codec pretraining used only the official training partition. As the datasets differ in acquisition protocols and release formats, we follow the official preprocessing pipeline published with each corpus. We apply across-channel normalization, preserving relative amplitudes across electrodes while bringing signals from different datasets onto a comparable scale range. For more details on the datasets, refer to Appendix~\ref{app:corpus}.

\subsection{Downstream Tasks}
\label{sec:downstreamtasks}

To assess the generalizability of \textsc{MyoCodec} across various applications,
we use \textsc{MyoCodec} as the EMG featurizer for four representative downstream tasks.
In \emph{emg2qwerty}, the task is to predict typed characters from wrist EMG signals, and \emph{emg2pose} estimates hand poses.
In \emph{emg2speech}, the aim is to decode speech waveforms from orofacial EMG signals, whereas \emph{speech2emg} addresses the reverse mapping, reconstructing EMG signals from speech waveforms.

For each task, we replace the original EMG representation module with \textsc{MyoCodec} while preserving the remainder of the architecture, apart from minor task-specific modifications.
We compare against TinyMyo~\citep{fasulo2025tinymyo}, BioCodec~\citep{avramidis2025biocodec}, and the corresponding task-specific EMG encoder or decoder from the original work.
Depending on the task, we use either the quantized codec output or continuous pre-quantization features, exploring the utility of different levels in the \textsc{MyoCodec} representation hierarchy.
All downstream tasks are trained causally unless stated otherwise. For controlled comparison, some task-specific data augmentation or perturbation is omitted for some downstream tasks. Full implementation details are provided in Appendix~\ref{app:downstream}.

\subsection{Evaluation}
Our evaluation on \textsc{MyoCodec} focuses on two aspects: intrinsic codec quality and application to downstream tasks.
\paragraph{Intrinsic codec quality.}
We assess reconstruction fidelity using scale-invariant signal-to-distortion ratio (SI-SDR)~\citep{leroux2019sdr}, signal-to-noise ratio (SNR), waveform-level Pearson correlation, and envelope correlation. We additionally evaluate the quality of the quantized representation using the utilization rate and perplexity of each codebook.
\paragraph{Downstream tasks.}
To assess generalizability, we evaluate performance on each downstream task using its commonly used evaluation metrics.
For \textit{emg2qwerty}, we report character error rate (CER).
For \textit{emg2pose}, we report mean joint-angle error with mean landmark and fingertip distances.
For \textit{emg2speech}, we report character, word, and phoneme error rates (CER, WER, and PER), phone feature error rate (PFER), and UTMOS~\citep{utmos}. PFER measures phonetic errors while accounting for similarity in phonological features, whereas UTMOS provides an automatic estimate of perceived speech quality.
For \textit{speech2emg}, we report envelope correlation of synthesized EMG signals.
Complete evaluation protocols and implementation details are provided in Appendix~\ref{app:downstream}.

\section{Results and Discussion}
\label{sec:results}

\begin{table}[tb]
\centering
\begin{threeparttable}
\caption{\textbf{emg2qwerty.} Character error rate (CER) is
reported with and without language-model beam search, across \emph{seen} and \emph{unseen} subjects. The best and second-best are \textbf{bolded} and \underline{underlined}, respectively.}
\label{tab:frontends}
\small
\setlength{\tabcolsep}{4pt}
\begin{tabular}{lrrrrrr}
\toprule
\multirow{3}{*}{EMG Representation} & \multirow{3}{*}{Hz} & \multirow{3}{*}{bitrate/ch}
  & \multicolumn{4}{c}{CER (\%)\,$\downarrow$} \\
\cmidrule(lr){4-7}
 & & & \multicolumn{2}{c}{greedy} & \multicolumn{2}{c}{LM beam} \\
\cmidrule(lr){4-5}\cmidrule(lr){6-7}
 & & & seen & unseen & seen & unseen \\
\midrule
Log-spectrogram \citep{sivakumar2024emg2qwerty} & 125.0 & $\le$132k$^\dagger$ & 24.02 & 52.98 & 16.86 & 48.57 \\
TinyMyo \citep{fasulo2025tinymyo} & 100.0 & $\le$614k$^\dagger$ & 21.16 & \textbf{50.02} & 14.48 & \textbf{44.55} \\
BioCodec \citep{avramidis2025biocodec} & 27.8 & 1.3k & \underline{18.52} & 51.87 & \underline{13.07} & 47.78 \\
\textsc{MyoCodec} (Ours) & 50.0 & 2.4k & \textbf{18.43} & \underline{50.47} & \textbf{12.74} & \underline{45.77} \\
\bottomrule
\end{tabular}
\begin{tablenotes}[flushleft]\footnotesize
\item[$\dagger$] Rate assumes \texttt{float32}.
\end{tablenotes}
\end{threeparttable}
\end{table}

\begin{figure}[t]
\centering
\includegraphics[width=0.98\textwidth]{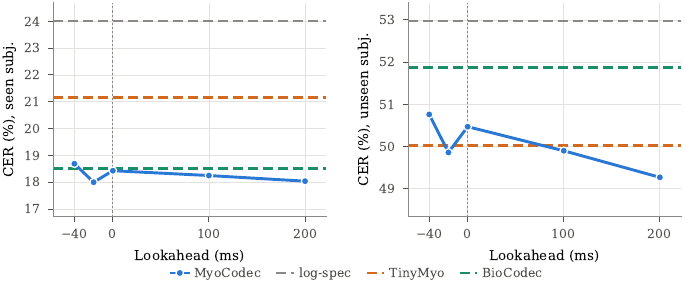}
\caption{\textbf{Effect of lookahead on emg2qwerty.} Decoding performance improves at a negative lookahead of 
\(-20\,\mathrm{ms}\), decreases slightly at zero lookahead, and then gradually improves
with increasing positive lookahead.}
\label{fig:lookahead}
\end{figure}

\subsection{Intrinsic Codec Quality}

We evaluate the intrinsic quality of \textsc{MyoCodec} along two dimensions: reconstruction fidelity and codebook usage. These analyses assess whether the codec preserves relevant signal information while learning a well-utilized discrete representation.

\paragraph{Signal reconstruction.}
\label{sec:recon}
As shown in Table~\ref{tab:recon}, \textsc{MyoCodec} achieves the best reconstruction performance across all reported metrics. \textsc{MyoCodec} already outperforms both BioCodec and TinyMyo on all four metrics when only two codebooks (\SI{800}{bits/s}) are used. The full codebook ablation results are shown in Figure~\ref{fig:rd} in Appendix~\ref{app:bitrate}. As depicted in Figure~\ref{fig:recon}, \textsc{MyoCodec} reconstructs the target EMG waveform better than TinyMyo and BioCodec, including transient peaks that the other models often miss.

\paragraph{Codebook usage.}
As shown in Table~\ref{tab:cbusage}, \textsc{MyoCodec} achieves full utilization of all six codebooks, indicating no representational collapse. Perplexity increases across successive RVQ stages, consistent with a coarse-to-fine decomposition in which the first stage captures the dominant signal structure and later stages encode progressively finer residual details.

\subsection{Downstream tasks}

\paragraph{emg2qwerty.}
\label{sec:emg2qwerty}

As shown in Table~\ref{tab:frontends}, \textsc{MyoCodec} achieves the best performance on seen subjects and second-best on unseen subjects, while operating at only \SI{2400}{bits/s} per channel. We also explore how different temporal lookaheads influence the decoding performance (Figure~\ref{fig:lookahead}). Decoding performance improves slightly at \SI{-20}{ms}, worsens at zero lookahead, and then gradually improves as additional future context is provided. One possible explanation is that muscle activation precedes the registered keystroke, whereas mechanical vibrations at key contact may introduce artifacts into the EMG signal; further analysis is required for more clarification, though. Note that, for controlled comparison, we omit the EMG augmentation methods used in the original \emph{emg2qwerty} framework \citep{sivakumar2024emg2qwerty}, including SpecAugment~\citep{park2019specaugment} and temporal jittering.

\paragraph{emg2pose.}
\label{sec:pose}

\begin{table}[t]
\centering
\caption{\textbf{emg2pose.} Mean joint-angle error (degrees) and mean landmark and fingertip distances (mm).}
\label{tab:pose}
\small
\begin{tabular}{lrrrr}
\toprule
\multirow{2}{*}{EMG encoder} & \multicolumn{2}{c}{joint-angle error ($^\circ$)\,$\downarrow$}
  & \multirow{2}{*}{land.\ (mm)\,$\downarrow$} & \multirow{2}{*}{fingertip (mm)\,$\downarrow$} \\
\cmidrule(lr){2-3}
 & val & test & & \\
\midrule
Log-spectrogram \citep{salter2024emg2pose} & 13.13 & 15.23 & 20.76 & 35.07 \\
TinyMyo \citep{fasulo2025tinymyo} & 13.15 & 14.83 & 20.13 & 33.97 \\
BioCodec \citep{avramidis2025biocodec} & 13.11 & 14.75 & 20.12 & 34.00 \\
\textsc{MyoCodec} (Ours) & \textbf{12.75} & \textbf{14.24} & \textbf{19.04} & \textbf{32.02} \\
\bottomrule
\end{tabular}
\end{table}

As shown in Table~\ref{tab:pose}, \textsc{MyoCodec} achieves the best performance across all three metrics: joint-angle error, landmark distance, and fingertip distance. All models use the same LSTM-based pose decoder from \citet{salter2024emg2pose} and only the input EMG representation differs.

\begin{figure}[tb]
\centering
\includegraphics[width=0.92\textwidth]{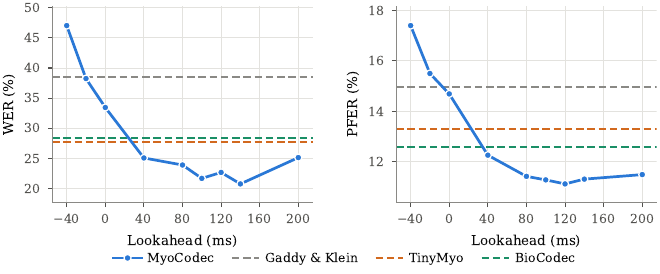}
\caption{\textbf{emg2speech with different lookahead.} The best performance is achieved when lookahead between \SI{100}{ms}--\SI{140}{ms} is applied.}
\label{fig:e2s-lookahead}
\end{figure}

\paragraph{emg2speech.}
\label{sec:e2s}
As shown in Table~\ref{tab:e2s}, \textsc{MyoCodec} outperforms all baselines across every evaluated metric, including CER, WER, PER, PFER, and UTMOS. Each downstream model is causally trained with \SI{100}{ms} lookahead, where the prediction target is JHCodec speech features~\citep{lee2026ssrr}. We use a retrained version of BioCodec specifically for this task, as its publicly released model weights were trained on band-pass-filtered EMG that excludes frequency content important for speech-related muscle activity. As shown in Figure~\ref{fig:e2s-lookahead}, the best performance is observed when lookahead between \SI{100}{ms}--\SI{140}{ms} is applied. Note that, to isolate the effect of the EMG representation, we apply no EMG augmentation or perturbation, including temporal jittering.
Speech samples are available here.\footnote{\url{https://lee-jhwn.github.io/myocodec}}

\begin{table}[tb]
\centering
\begin{threeparttable}
\caption{\textbf{emg2speech results.} We report character, word, and phoneme error rates (CER, WER, and PER), phone feature error rate (PFER), and UTMOS.
All of the downstream models are causally trained with \SI{100}{ms} of lookahead.}
\label{tab:e2s}
\setlength{\tabcolsep}{5pt}
\begin{tabular}{lrrrrr}
\toprule
EMG Encoder & CER\,$\downarrow$ & WER\,$\downarrow$ & PER\,$\downarrow$ & PFER\,$\downarrow$ & UTMOS\,$\uparrow$ \\
\midrule
Ground truth & 1.43 & 3.76 & 7.13 & 4.19 & 3.41 \\
\midrule
Gaddy \& Klein \citep{gaddy2021improved} & 22.95 & 38.49 & 31.62 & 14.98 & 1.90 \\
TinyMyo \citep{fasulo2025tinymyo} & 16.39 & 27.77 & 26.47 & 13.29 & 1.91 \\
BioCodec$^\dagger$ & 16.56 & 28.33 & 24.63 & 12.59 & 1.93 \\
\textsc{MyoCodec} (Ours) & \textbf{12.15} & \textbf{21.12} & \textbf{21.61} & \textbf{11.54} & \textbf{2.00} \\
\bottomrule
\end{tabular}
\begin{tablenotes}[flushleft]\footnotesize
\item[$\dagger$] We used a retrained version of BioCodec, as the preprocessing used for its released checkpoint removes the frequency range crucial for speech-related information in EMG signals.
\end{tablenotes}
\end{threeparttable}
\end{table}

\paragraph{speech2emg.}
\label{sec:speech2emg}
Table~\ref{tab:s2e} compares \textsc{MyoCodec} with baseline systems under causal, limited-lookahead, and offline inference. Following \citet{scheck2023stegan}, we use envelope correlation coefficient (Env.\ CC) as a proxy metric to assess the quality of generated EMG signals, evaluated with \SI{20}{ms} and \SI{50}{ms} smoothing windows. Non-causally trained STE-GAN~\citep{scheck2023stegan} achieves the best performance, while causally trained \textsc{MyoCodec} closely matches \SI{120}{ms} of lookahead. Note that STE-GAN and TinyMyo are trained non-causally, while BioCodec and \textsc{MyoCodec} are trained causally.

\begin{table}[tb]
\centering
\caption{\textbf{speech2emg.} Envelope correlation coefficients
(Env.\ CC) with 95\% confidence intervals are reported using \SI{20}{ms} and \SI{50}{ms}
smoothing windows. We compare offline and causal systems with different lookahead. The best and second-best are \textbf{bolded} and \underline{underlined}, respectively.}
\label{tab:s2e}
\small
\setlength{\tabcolsep}{4.5pt}
\begin{tabular}{llcc}
\toprule
 & & \multicolumn{2}{c}{Env.\ CC\,$\uparrow$} \\
\cmidrule(lr){3-4}
System & lookahead & \SI{50}{ms} window & \SI{20}{ms} window \\
\midrule
STE-GAN \citep{scheck2023stegan} & $\infty$ (offline) & $\mathbf{0.705}$\,\ci{0.014} & $\mathbf{0.578}$\,\ci{0.012} \\
TinyMyo \citep{fasulo2025tinymyo} & $\infty$ (offline) & $0.577$\,\ci{0.011} & $0.474$\,\ci{0.010} \\
\midrule
\multirow{2}{*}{BioCodec \citep{avramidis2025biocodec}}
 & \SI{0}{ms} & $0.511$\,\ci{0.010} & $0.434$\,\ci{0.008} \\
 & \SI{144}{ms} & $0.600$\,\ci{0.009} & $0.522$\,\ci{0.008} \\
\midrule
\multirow{2}{*}{\textsc{MyoCodec} (Ours)}
 & \SI{0}{ms} & $0.587$\,\ci{0.010} & $0.503$\,\ci{0.009} \\
 & \SI{120}{ms} & $\underline{0.655}$\,\ci{0.010} & $\mathbf{0.578}$\,\ci{0.009} \\
\bottomrule
\end{tabular}
\end{table}

\paragraph{Streaming inference.}
\label{sec:cost}
We measure inference speed on a single \texttt{RTX PRO 6000 Blackwell} GPU. Table~\ref{tab:cost} compares inference speed for each codec on two bases: per-second chunks and per-frame streaming. On a per-second basis, \textsc{MyoCodec} requires \SI{1.474}{ms} for encoding and decoding, compared with \SI{2.032}{ms} for BioCodec. For frame-level streaming inference, \textsc{MyoCodec} maintains a state cache and processes each \SI{20}{ms} frame using compiled \texttt{CUDA Graphs}. Encoding and decoding require \SI{0.253}{ms} and \SI{0.229}{ms}, respectively, for a total of \SI{0.482}{ms} per frame, corresponding to \(41\times\) real-time throughput. On a single CPU thread (Intel Xeon Platinum 8559C), streaming inference compute times are 3.27\,ms for encoding and 2.91\,ms for decoding, 6.18\,ms in total, corresponding to a real-time throughput of $3.24\times$.

\subsection{Potential Future Applications}

\textsc{MyoCodec}'s compact size, causal operation, and discrete output at low bitrate make it promising for on-device streaming. Processing EMG input directly on a wearable could enable low-latency interaction, reduce transmission bandwidth, and limit the need to transmit raw biosignals.
As EMG signals constitute sensitive physiological data, local processing could also improve privacy by keeping raw signals on the device.
These properties could broaden the practical and secure use of wearable EMG devices in assistive communication, gesture and prosthetic control, and hands-free input.

Beyond on-device deployment, encoding continuous EMG signals into low-bitrate discrete tokens provides a potential direct integration to modern language and speech-language models. As modern speech and language modeling approaches mostly operate through next-token prediction over discrete representations, \textsc{MyoCodec} makes EMG signals more compatible with similar architectures and training objectives. This creates a path toward interactive multimodal systems that condition directly on muscle activity to generate text, speech, or actions without intermediate steps, potentially reducing the latency, information loss, and error propagation associated with cascaded pipelines.

The high reconstruction fidelity achieved by \textsc{MyoCodec}'s decoder suggests a complementary application in synthetic EMG generation. Generative models operating on \textsc{MyoCodec} tokens could synthesize EMG signals from speech, motion, or task descriptors, providing synthetic training data when collecting large-scale labeled EMG is costly or impractical. Such signals could augment data-scarce training sets and support neuromuscular digital-human models that simulate task- and subject-dependent muscle activity. Realizing these applications, however, will require further validation of physiological realism, cross-channel consistency, and downstream utility.

\subsection{Limitations}
\label{sec:limitations}

The publicly available EMG datasets are highly imbalanced to a couple of main sources, \textit{emg2qwerty}~\citep{sivakumar2024emg2qwerty} and \textit{emg2pose}~\citep{salter2024emg2pose}, accounting for \(71\%\) of the collected data, whereas speech-related EMG data contribute only \(1.9\%\) (Appendix~\ref{app:corpus}). Such concentration could potentially limit uniform generalization across the broader range of EMG applications, sensor configurations, and recording conditions.

Our downstream comparisons also prioritize experimental control over task-specific optimization. For each downstream task, we change the EMG representation primarily while retaining the rest of the architecture. However, most downstream task models were originally designed for their original continuous EMG feature inputs and may not fully exploit discrete codec tokens. The reported results should therefore be interpreted as measuring the \textsc{MyoCodec}'s plug-in compatibility, rather than the maximum performance attainable with architectures designed specifically for such EMG representation.

\begin{table}[t]
\centering
\begin{threeparttable}
\caption{\textbf{Inference speed.} All measurements are obtained on a single \texttt{NVIDIA RTX PRO 6000 Blackwell} GPU. Per-second speed is measured using \SI{5}{s} input windows and normalized by signal duration, while \textsc{MyoCodec}'s per-frame speed is measured using its compiled \texttt{CUDA Graph} implementation.}
\label{tab:cost}
\small
\begin{tabular}{llrrrrrr}
\toprule
 & & \multicolumn{3}{c}{Compute time (ms) $\downarrow$} & \multicolumn{3}{c}{Real-time throughput ($\times$) $\uparrow$} \\
\cmidrule(lr){3-5}\cmidrule(lr){6-8}
System & basis & Encoder & Decoder & Total & Encoder & Decoder & Total \\
\midrule
\multirow{2}{*}{\textsc{MyoCodec} (Ours)}
 & per second & 0.782 & 0.692 & 1.474 & 1279 & 1445 & 678 \\
 & per frame & 0.253 & 0.229 & 0.482 & 79 & 87 & 41 \\
\midrule
\multirow{2}{*}{BioCodec}
 & per second & 1.091 & 0.941 & 2.032 & 917 & 1062 & 492 \\
 & per frame & \multicolumn{6}{c}{\cellcolor{igray!45}N/A$^\dagger$} \\
\bottomrule
\end{tabular}
\begin{tablenotes}[flushleft]\footnotesize
\item[$\dagger$] The official BioCodec release provides neither a state cache nor a frame-wise streaming interface.
\end{tablenotes}
\end{threeparttable}
\end{table}

\section{Conclusion}

We introduce \textsc{MyoCodec}, a streaming neural codec that encodes EMG signals at \SI{50}{Hz} and \SI{2400}{bits/s} using causal Transformer blocks and residual vector quantization. Trained on twelve public corpora, \textsc{MyoCodec} achieves the strongest reconstruction performance among the compared models. Its EMG representations can generalize across multiple downstream tasks, such as \textit{emg2qwerty}, \textit{emg2pose}, \textit{emg2speech}, and \textit{speech2emg}.
This provides a practical foundation for on-device wearable systems and, more broadly, for direct integration of muscle activity with modern language and speech-language models.

\subsection*{Reproducibility Statement}
To support reproducibility, we will publicly release the complete codebase and model weights upon publication.

\subsection*{Ethics statement}
This work uses only publicly released EMG datasets, collected and distributed by
their original authors under their own consent and IRB arrangements. No additional human-subject data were collected.

\subsection*{AI use statement}
We used AI for literature summarization, experimental implementation assistance, and refining writing. The authors reviewed AI-assisted code implementation and verified its outputs. The research ideas, experimental design, interpretation of results, and scientific claims are from the authors. The authors take full responsibility for the final content of this work.



\bibliography{iclr2027_conference}
\bibliographystyle{iclr2027_conference}

\appendix

\section{Related Work}
\label{app:related}

\paragraph{Biosignal foundation models.}
A parallel line of work pretrains large encoders on physiological signals and transfers their continuous representations to downstream tasks. LaBraM \citep{jiang2024labram}, for example, pretrains a masked-prediction Transformer on thousands of hours of EEG data drawn from around dozens of different datasets. It uses a vector-quantized spectral tokenizer to construct pretraining targets, transferring the resulting encoder to abnormality detection, event classification, and emotion recognition.
For EMG, \citet{kaifosh2025neuromotor} train a unified wrist-worn decoder across thousands of participants and demonstrate generalization to previously unseen users without person-specific calibration. This provides strong evidence that pretrained EMG representations can transfer across individuals. TinyMyo \citep{fasulo2025tinymyo} is a compact EMG foundation model built for edge deployment. It embeds non-overlapping 20-sample patches of each channel, giving a frame rate of \SI{100}{Hz}, and carries a learned bank of 16 channel embeddings so that a single model serves different electrode layouts. At roughly \SI{3.6}{M} parameters, it is one of the smallest of these encoders, outputting 192-dimensional continuous embeddings. We use TinyMyo as a baseline representation on all four downstream tasks.

\paragraph{Neural audio codecs.}
Discrete neural audio codecs typically combine an encoder, RVQ, and a decoder trained with reconstruction and adversarial objectives.
SoundStream \citep{zeghidour2021soundstream} introduced the RVQ-GAN formulation, 
EnCodec \citep{defossez2022encodec} incorporated multi-scale spectral discrimination and streaming inference, and DAC \citep{kumar2023dac} improved codebook utilization through factorized and \(L_2\)-normalized codes. 
Mimi \citep{defossez2024moshi} further distilled phonetic information into the first RVQ stage, producing tokens useful for language modeling as well as waveform reconstruction.
TS3-Codec \citep{ts3codec} adopts a fully causal Transformer decoder for neural audio codecs. 
JHCodec \citep{lee2026ssrr} extends this framework with a self-supervised representation reconstruction loss to enable stable streaming without lookahead.

\paragraph{BioCodec.}
BioCodec~\citep{avramidis2025biocodec} is a general-purpose neural codec for biosignal tokenization and therefore serves as our primary codec baseline. It follows an EnCodec-style convolutional architecture comprising a causal SEANet encoder and a six-stage RVQ with 256 entries per stage. In its EMG configuration, BioCodec operates at \(\SI{27.78}{Hz}\) and \SI{1333}{bits/s} per channel. The EMG model was pretrained on \textit{emg2qwerty}~\citep{sivakumar2024emg2qwerty} and evaluated primarily as a tokenizer for downstream gesture classification.
BioCodec~\citep{avramidis2025biocodec} is a general-purpose neural codec for biosignal tokenization and therefore serves as our primary codec baseline. It follows an EnCodec-style convolutional architecture comprising a causal SEANet encoder and a six-stage RVQ with 256 entries per stage. In its EMG configuration, BioCodec operates at \(\SI{27.78}{Hz}\) and \SI{1333}{bits/s} per channel. The EMG model was pretrained on \textit{emg2qwerty}~\citep{sivakumar2024emg2qwerty} and evaluated primarily as a tokenizer for downstream gesture classification.

\paragraph{emg2qwerty.}
The task is to decode wrist-worn EMG into the keystrokes a person types~\citep{sivakumar2024emg2qwerty}.
Its published front end is a log-spectrogram
followed by a rotation-invariant band MLP and a time-depth-separable convolutional
encoder trained with CTC, and its headline result is that a generic model trained across
participants transfers to held-out ones.

\paragraph{emg2pose.}
The target is to decode the joint angles of the hand, with
motion capture providing the reference. \citet{salter2024emg2pose} pair wrist EMG with an
optical hand-tracking rig over 193 participants and release both the corpus and a
recurrent decoder.

\paragraph{emg2speech.}
Articulatory EMG recorded from the face and neck carries enough information to decode
speech, including when the speech is silently mouthed.
\citet{schultz2010coarticulation} established continuous recognition from this signal and
showed that modeling coarticulation is what makes it tractable, and
\citet{janke2017emgtospeech} went further by generating a speech waveform directly from
the EMG rather than passing through text. \citet{gaddy2020digital, gaddy2021improved} released a single-speaker \textit{emg2speech} corpus with both vocalized and silent speech.
\citet{meltzner2018semg} moved the setting toward practical use, developing
EMG sensor arrays and recognition algorithms for silent speech with a vocabulary large
enough to be useful rather than demonstrative. \citet{gowda2025emg2speech} extend the
direction to a multi-speaker corpus, including a participant with ALS. A related
line explores what information is encoded in the signals.
\citet{lee2025articulatory} predict articulatory features directly from surface EMG during
speech production, which is used as the intermediate representation for speech synthesis.
Taking that further, \citet{lee2026simultaneous} explore acquiring real-time MRI, EEG, and EMG simultaneously
during speech, which places EMG explicitly between the neural command and the articulator
motion it produces. \citet{pistrosch2026affect} decode affect from surface EMG in both
phonated and silent production. \citet{hwang2026fewer} study
how far the electrode count can be reduced, finding that the best subsets are those whose
channels are complementary rather than individually strongest.

\paragraph{speech2emg.}
This is the reverse task of \textit{emg2speech}, namely synthesizing EMG from speech input.
\citet{scheck2023stegan} introduce STE-GAN, an adversarial model that generates
articulatory EMG conditioned on speech features.

\section{Bitrate Comparison}
\label{app:bitrate}
\begin{figure}[t]
\centering
\includegraphics[width=\textwidth]{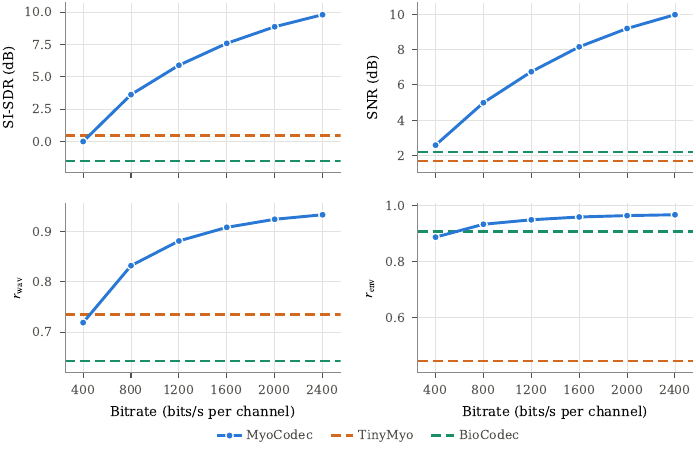}
\caption{\textbf{Reconstruction quality across various bitrates.} \textsc{MyoCodec} outperforms TinyMyo and BioCodec on reconstruction quality using only two codebooks.}
\label{fig:rd}
\end{figure}

As the quantizer is residual, it allows to use only the first \(N_q\) codebooks, reducing the bitrate in increments of \SI{400}{bits/s}. Figure~\ref{fig:rd} shows the bitrate--quality trade-off for \(N_q\in\{1,\ldots,6\}\). Using only two codebooks (\SI{800}{bits/s}, approximately 60\% of BioCodec's \SI{1333}{bits/s}), \textsc{MyoCodec} outperforms BioCodec across all four reconstruction metrics. With only one codebook (\SI{400}{bits/s}), \textsc{MyoCodec} continues to lead in SI-SDR, SNR, and \(r_{\mathrm{wav}}\), while trailing in \(r_{\mathrm{env}}\).

\section{Detailed Implementation of \textsc{MyoCodec}}
\label{app:codec}

\subsection{Architecture}

\paragraph{Encoder.}
All EMG inputs are sampled at \SI{2}{kHz} and patchified into non-overlapping frames of
\(F=40\) samples, corresponding to a frame duration of \SI{20}{ms} and a token rate of \SI{50}{Hz}.
We treat each channel independently, as the datasets use different electrode configurations, supporting arbitrary electrode configurations. Each frame is projected linearly from
40 samples to a model width of \(d=256\).
The encoder comprises eight pre-norm Transformer layers, each with eight attention heads
of dimension 32, an MLP hidden dimension of 1024, and no dropout. We use rotary positional
embeddings \citep{su2021roformer} and FlashAttention
\citep{dao2022flashattention} for all attention operations. Self-attention is causal and
restricted to a 64-frame sliding window. 
During streaming inference, each layer retains a 64-frame key--value
cache, keeping the per-frame computation independent of sequence length. The encoder size is approximately \(6.3\,\mathrm{M}\) parameters.

\paragraph{Quantizer.}
The encoder output is projected from 256 to 32 dimensions before quantization. We use a
residual vector quantizer with \(N_q=6\) stages, each containing \(K=256\) entries of
dimension 32. A learned projection maps the quantized representation back to 256
dimensions before fed into the decoder.
Codebook entries are initialized from \(\mathcal{N}(0,1)\), giving an expected norm of
approximately \(\sqrt{32}\). Assignment counts are tracked separately for each entry
using an exponential moving average with decay \(0.99\). Every 250 optimization steps,
entries whose averaged count falls below \(0.9\) are reinitialized from encoder outputs
sampled from the current batch, and their count is reset to \(1.8\). This mechanism
provides a grace period after reinitialization, maintains full utilization of all six
codebooks, and helps prevent codebook collapse.

\paragraph{Decoder.}
The decoder mirrors the encoder, using eight pre-norm causal Transformer layers with the
same width, attention configuration, MLP dimension, rotary embeddings, and 64-frame
sliding window. It takes the dequantized 256-dimensional sequence as input and applies a final
linear projection to reconstruct \(F=40\) waveform samples at each frame. During
streaming, the decoder maintains the same per-layer key--value cache as the encoder. It
contains approximately \(6.3\,\mathrm{M}\) parameters; including the quantizer and
projection layers, the complete codec contains \(12.72\,\mathrm{M}\) parameters.

\paragraph{Discriminators.}
Adversarial training uses two complementary discriminator branches. The multi-period
discriminator reshapes the waveform using periods
\(\{2,3,5,7,11\}\), with 8 base filters increasing to at most 128. The multi-scale STFT
discriminator uses FFT and window sizes \(\{64,128,256,512\}\), hop lengths
\(\{17,31,67,131\}\), and 16 base filters increasing to at most 128. The discriminators are used only
during Phase~2 and are discarded at inference.

\subsection{Training Objectives}

\paragraph{Phase~1.}
Phase~1 optimizes the codec objective in Equation~\ref{eq:gen}, which combines a
time-domain Huber loss, multi-scale spectral reconstruction, vector-quantization loss,
and commitment loss. 
The spectral objective is averaged over FFT sizes \(\{64,128,256\}\), corresponding to
\SIrange{32}{128}{ms} at \SI{2}{kHz}. Each scale uses a Hann window with hop size \(n/4\)
and combines log-magnitude error, spectral convergence, and phase error, with the phase
term weighted by \(0.1\) as defined in Equation~\ref{eq:spec}. Magnitudes are floored at
\(10^{-7}\) before taking the logarithm, and phase is compared using principal-value
angles without unwrapping. Phase~1 runs for \num{100}k steps on the Muon optimizer~\citep{jordan2024muon} with a batch
size \num{1024} and no discriminator.

\paragraph{Huber loss.}
The time-domain reconstruction term is the mean Huber penalty between the target \(x\)
and reconstruction \(\hat{x}\):
\[
\mathcal{L}_{\mathrm{Huber}}(x,\hat{x})
=
\frac{1}{T}\sum_{t=1}^{T}
h_{\beta}\!\left(\hat{x}_t-x_t\right),
\qquad
h_{\beta}(d)
=
\begin{cases}
\tfrac{1}{2}d^2, & |d|<\beta,\\[2pt]
\beta\left(|d|-\tfrac{1}{2}\beta\right), & |d|\geq\beta.
\end{cases}
\]
where \(\beta=1\). The quadratic region provides smooth gradients for small
reconstruction errors, while the linear region limits the influence of isolated
high-amplitude bursts and motion artifacts.

\paragraph{Vector-quantization losses.}
Let \(r^{(1)}=z\) denote the projected encoder output, \(e^{(i)}\) the nearest entry
selected from codebook \(i\), and \(\operatorname{sg}[\cdot]\) the stop-gradient
operator. Successive RVQ stages operate on the residual
\[
r^{(i+1)}
=
r^{(i)}-\operatorname{sg}\!\left[e^{(i)}\right].
\]
The vector-quantization and commitment losses are summed over all \(N_q=6\) stages:
\[
\mathcal{L}_{\mathrm{VQ}}
=
\sum_{i=1}^{N_q}
\left\|
\operatorname{sg}\!\left[r^{(i)}\right]-e^{(i)}
\right\|^2,
\qquad
\mathcal{L}_{\mathrm{commit}}
=
\sum_{i=1}^{N_q}
\left\|
r^{(i)}-\operatorname{sg}\!\left[e^{(i)}\right]
\right\|^2.
\]
Each squared norm is averaged over the 32 code dimensions and the batch and time axes.
The two losses differ in their gradient paths: \(\mathcal{L}_{\mathrm{VQ}}\) updates the
selected codebook entries, whereas \(\mathcal{L}_{\mathrm{commit}}\) updates the encoder
and discourages its outputs from drifting away from the codebook. We assign the
commitment term a weight of \(0.25\), compared with unit weight for the VQ loss.

Let \(q=\sum_{i=1}^{N_q}e^{(i)}\) denote the quantized representation. As nearest-neighbor assignment is non-differentiable, the decoder receives the straight-through
estimate
\[
z+\operatorname{sg}[q-z],
\]
which equals \(q\) during the forward pass but has an identity gradient with respect to
\(z\). Detaching each selected entry in the residual recursion also prevents later RVQ
stages from propagating gradients through earlier codebooks.

\paragraph{Phase 2.}
Phase~2 retains the complete Phase~1 objective and adds the adversarial and
feature-matching terms in Equation~\ref{eq:adv}. Their target weights are
\[
\lambda_{\mathrm{adv}}=0.1,
\qquad
\lambda_{\mathrm{FM}}=2.0.
\]
Both weights are increased linearly from zero over the first \num{50000} steps of
Phase~2, providing a gradual transition to adversarial training. The discriminators are
optimized with AdamW using a learning rate of \(10^{-4}\), weight decay of \(10^{-4}\),
and gradient clipping at \(1.0\). As the discriminators dominate memory usage, the
batch size is reduced from \num{1024} to \num{192}. Phase~2 runs for an additional
\num{100000} steps, giving \num{200000} optimization steps in total.

\section{Downstream implementation}
\label{app:downstream}

\paragraph{emg2qwerty.}
The downstream model only differs in inputted EMG features.
For \textsc{MyoCodec} and BioCodec, the downstream model receives the quantized RVQ output, while TinyMyo takes continuous encoder features from its encoder.
We retain the multi-band rotation-invariant MLP and
CTC output layer from \citet{sivakumar2024emg2qwerty}.
The MLP has one hidden layer of width
384 and produces a 768-dimensional representation, followed by a
\texttt{Linear(768, 99)} classifier.
The temporal encoder is a four-layer causal
Transformer with input and output projections of \(768\rightarrow256\) and
\(256\rightarrow768\), respectively. Each layer uses four attention heads of dimension
64, an MLP hidden dimension of 1536, and dropout \(0.1\).
Beam-search decoding follows \citet{sivakumar2024emg2qwerty}, using a beam width of 50,
a language-model weight of \(2.0\), an insertion bonus of \(2.0\), and a 6-gram character
language model trained on WikiText-103. To isolate the effects of the input representation
and temporal lookahead, we omit EMG augmentation. In particular,
SpecAugment~\citep{park2019specaugment} is not applicable to the non-spectrogram \textsc{MyoCodec} representation,
and temporal jittering is excluded because it would confound the lookahead analysis.

\paragraph{emg2pose.}
The codec replaces the \texttt{TdsNetwork} featurizer~\citep{salter2024emg2pose}, a causal strided convolutional network that maps 16-channel EMG sampled at \SI{2}{kHz} to 64-dimensional features at \SI{50}{Hz} with an overall stride of 40.
For all three front ends the downstream model receives the encoder output, the hidden state before quantization. Multi-head attention mixes information across the 16 electrodes within each frame without introducing additional lookahead, and a two-layer projection with hidden width 256 maps the result to a 128-dimensional feature vector. All subsequent components are retained from the published \textit{emg2pose} model \citep{salter2024emg2pose}, including the velocity-integrating LSTM decoder, whose 148-dimensional input concatenates the projected EMG features with the 20 joint angles from the previous predicted state.

\paragraph{emg2speech.}
The three pretrained front ends are frozen and provide continuous features. The prediction target is the continuous 1024-dimensional pre-quantization latent of JHCodec. For the three pretrained front ends, a three-layer bidirectional channel Transformer mixes the eight electrodes within each frame at width 256 using four attention heads and dropout \(0.2\), followed by a linear projection to 512 dimensions. A causal temporal Transformer then models the sequence using eight layers, four 128-dimensional attention heads, an MLP hidden dimension of 2048, dropout \(0.2\), rotary embeddings, and a 128-frame sliding window. The output head consists of \(512\rightarrow1024\), SiLU, and \(1024\rightarrow1024\) layers, and an auxiliary \(512\rightarrow48\) phoneme classifier contributes to the loss with weight \(0.5\). The JHCodec latent is predicted using an \(\ell_1\) loss. All components from the temporal Transformer onward are shared across the four systems. Training uses the vocalized set of the Gaddy and Klein corpus. Note that the Gaddy and Klein encoder is bidirectional, serving as an offline reference.

\paragraph{speech2emg.}
\emph{speech2emg} predicts EMG from speech. The input
is JHCodec's 1024-dimensional pre-quantization speech representation at
\SI{50}{Hz}. The continuous prediction target depends on the EMG encoder:
\textsc{MyoCodec} and BioCodec use the bottleneck projection, and TinyMyo uses its encoder output.
The predictor comprises a six-layer causal temporal Transformer (width 512,
eight heads) and a two-layer bidirectional channel Transformer over the eight
electrodes (width 256, four heads), followed by a linear regression head.
Dropout is \(0.1\). The objective combines mean-squared error with
STE-GAN's multi-time-domain loss \citep{scheck2023stegan}.

\paragraph{Evaluation metrics.} We report a separate set of evaluation metrics per considered task:
\begin{itemize}[leftmargin=1.2em,itemsep=3pt,topsep=2pt,parsep=0pt]

\item \emph{emg2qwerty.}
We report CER, which is measured using greedy decoding and the beam-search decoder with a 6-gram character language model, identical to \cite{sivakumar2024emg2qwerty}.

\item \emph{emg2pose.}
We report the mean absolute error of the 20 joint angles in degrees. We additionally report Euclidean landmark
and fingertip distances in millimetres after applying forward kinematics with the
published hand model. Landmark distance is averaged over all articulated landmarks, and
fingertip distance over the five fingertips. 

\item \emph{emg2speech.}
We report CER, WER, PER, and PFER, and UTMOS~\citep{utmos}. CER and WER are measured using Whisper large-v3~\citep{whisper}. PER and PFER are measured by a
wav2vec~2.0 phoneme recognizer~\citep{baevski2020wav2vec2,xu2022phoneme}. PFER assigns partial substitution
costs according to shared articulatory features, while UTMOS provides a proxy to perceptual quality.

\item \emph{speech2emg.}
We report the envelope correlation coefficient (Env.\ CC). Following STE-GAN~\citep{scheck2023stegan}, envelopes are obtained by rectifying each signal and
applying a moving-average filter.

\end{itemize}

\section{Pretraining corpus}
\label{app:corpus}

\begin{table}[h]
\centering
\begin{threeparttable}
\caption{The twelve public EMG corpora. Channel-hours count only the channels and recordings retained after filtering.}
\label{tab:corpus}
\small
\setlength{\tabcolsep}{4.5pt}
\begin{tabular}{lrrrl}
\toprule
Corpus & ch. & subj. & ch.-hours & content \\
\midrule
emg2qwerty \citep{sivakumar2024emg2qwerty} & 32 & 108 & 11{,}074 & typing \\
emg2pose \citep{salter2024emg2pose} & 16 & 193 & 5{,}925 & hand pose \\
Hyser \citep{jiang2021hyser} & 256 & 20 & 2{,}593 & finger force \\
putEMG \citep{kaczmarek2019putemg} & 24 & 44 & 1{,}045 & hand gestures \\
EMG-EPN-612 \citep{benalcazar2020emgepn} & 8 & 612 & 1{,}014 & hand gestures \\
MeganePro \citep{cognolato2020meganepro} & 12 & 45 & 610 & grasping \\
GRABMyo \citep{pradhan2022grabmyo} & 28 & 43 & 597 & hand gestures \\
Ninapro$^\ast$ & 12--14 & 82 & 503 & hand movements \\
emg2speech (ALS) \citep{gowda2025emg2speech} & 31 & 2 & 304 & speech (incl.\ ALS) \\
CSL-HDEMG \citep{amma2015hdemg} & 192 & 5 & 232 & finger gestures \\
emg2speech (Gaddy) \citep{gaddy2020digital} & 8 & 1 & 156 & speech \\
CapgMyo DB-a \citep{geng2016capgmyo} & 128 & 18 & 51 & hand gestures \\
\midrule
\textbf{Total} & & & \textbf{24{,}104} & \\
\bottomrule
\end{tabular}
\begin{tablenotes}[flushleft]\footnotesize
\item[$\ast$] Only DB2~\citep{atzori2014ninapro}, DB4~\citep{pizzolato2017ninapro}, DB6~\citep{palermo2017ninapro} and DB7~\citep{krasoulis2017ninapro} are used.
\end{tablenotes}
\end{threeparttable}
\end{table}

\end{document}